\documentclass[%
 reprint,
superscriptaddress,
 amsmath,amssymb,
 aps,
prb,
]{revtex4-1}
\usepackage{graphicx}
\usepackage{dcolumn}
\usepackage{bm}
\usepackage{txfonts}
\usepackage{bm}
\begin{document}

\title{Fractionally Quantized Berry's Phase in an Anisotropic Magnet on the Kagome Lattice}

\author{Tohru Kawarabayashi}
\affiliation{Department of Physics, Toho University,
Funabashi, 274-8510 Japan}

\author{Kota Ishii}
\affiliation{Department of Physics, Toho University,
Funabashi, 274-8510 Japan}

\author{Yasuhiro Hatsugai}
\affiliation{Department of Physics, University of Tsukuba, Tsukuba, 305-8571 Japan}


\begin{abstract}
A fractionally quantized Berry phase is examined numerically in an anisotropic spin-1/2 XXZ model on the Kagome lattice.
It is shown that the Berry phase has a fractionally quantized and non-zero value when anisotropy is increased, 
which is consistent with the emergence of the tripartite entangled plaquette phase discussed by 
J. Carrasquilla et al. [Phys. Rev. B96, 054405 (2017)].
\end{abstract}


\maketitle

It has been pointed out that the Berry phase\cite{Berry} associated with an adiabatic and cyclic change of a parameter in Hamiltonians can be adopted as a topological order 
parameter to characterize the quantum state for which the order parameters in the conventional sense do not exist\cite{Hatsugai1,MH,HM,KH,KMH}.  

In the recent study on an anisotropic magnet on the two-dimensional Kagome lattice\cite{CCM,CL,CKK}, a new phase called 
the tripartite entangled plaquette state has been found \cite{CCM} by using large-scale Monte Carlo simulations when the 
average magnetization is 1/3 of the saturation magnetization.
This new phase appears when anisotropy is introduced in the antiferromagnetic interactions between the $z$-components of the spins.  
It is different from the valence-bond solid phase and the ferromagnetic phase 
reported in the isotropic couplings \cite{DS,IWMSK}, 
it restores all the symmetries of the Hamiltonian and is gapped. More interestingly, 
it is adiabatically connected \cite{CCM} to the product state of the tripartite entangled 
three-qubit W states \cite{DVC} in the limit of the strong anisotropy.
This means that the ground state of the anisotropic Kagome magnet at the average magnetization of 1/3, 
is essentially a short-ranged entangled state, where 
the entanglement may be detected by the Berry phase.\cite{Hatsugai1,KH}
Therefore, in the present paper, to characterize the tripartite entangled plaquette state of the anisotropic Kagome magnet, we examine the possibility of the $Z_3$ Berry phase numerically.

\begin{figure}
\includegraphics[scale=0.3]{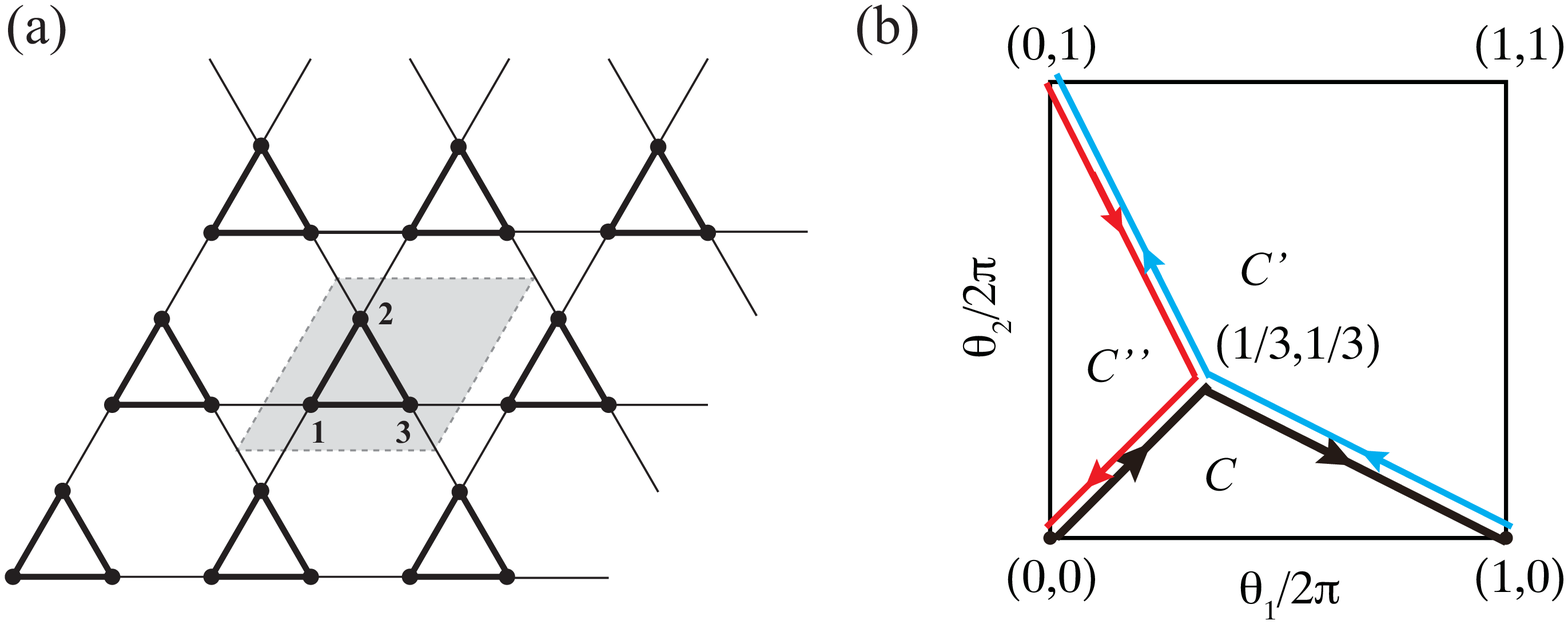}
\caption{(Color Online)
(a) The Kagome lattice. The unit cell is indicated by the shaded area. In the thick (thin) solid triangles, 
the antiferromagnetic coupling for the $z$-component is assumed to be  $J_z^u(J_z^d)$. 
(b) The path $C$ ($(0,0) \to (2\pi/3,2\pi/3) \to (2\pi, 0)$) of the parameter $\Theta=(\theta_1,\theta_2)$ to define the 
$Z_3$ Berry phase. Because of the three-fold symmetry of the lattice, the path $C$(black) is equivalent to the path $C'$($(2\pi,0) \to (2\pi/3,2\pi/3) \to (0, 2\pi))$ (blue)
and $C''$($(0,2\pi) \to (2\pi/3,2\pi/3) \to (0, 0))$(red).\cite{HM}
\label{fig1}
}
\end{figure}

We consider a spin-1/2 XXZ model on the Kagome lattice as shown in Fig. \ref{fig1}(a), 
where the unit cell is assumed to have three intra-spins forming the thick solid triangles (the up triangles). These intra-spins are coupled by thin lines to the spins in the neighboring unit cells, 
which form the down triangles. 
The Hamiltonian we adopt is described as\cite{CCM}
$$
 H = \sum_{<i,j>} -J_{xy}(S_i^xS_j^x + S_i^yS_j^y) + J_{z}^a S_i^z S_j^z, \quad a=u,d
$$
where $J_{xy}>0$ denotes the ferromagnetic coupling in the $x$-$y$ plane and $J_{z}^u >0$ denotes the 
antiferromagnetic coupling between $z$-component of intra-sites (the up triangles) 
in the unit cell of the Kagome lattice, whereas $J_{z}^d>0$ denotes the coupling between inter sites represented by the thin
solid lines (the down triangles). Here, the spin-1/2 operators at the lattice site $i$ are denoted 
by $\bm{S}_i = (S_i^x,S_i^y,S_i^z)$. The summation is performed over the nearest-neighbor pairs $<i,j>$ of the 
Kagome lattice. 

Different from the conventional definition of the Berry phase, here we consider the Berry phase 
for a cyclic change of local interactions in the lattice model.\cite{Hatsugai1} 
We consider a two dimensional parameter $\Theta = (\theta_1,\theta_2)$  
for a set of three intra-spins in a certain unit cell (the up triangle) \cite{HM} and 
modify the interactions among these three intra-spins as
\cite{Hatsugai1}
\begin{eqnarray*}
  H(\bm{S}_1,\bm{S}_2,\bm{S}_3) &= &-\frac{J_{xy}}{2}\bigg\{ (e^{i\theta_1}S_1^+S_2^- + e^{-i\theta_1}S_1^-S_2^+) \\
 & & + (e^{i\theta_2}S_2^+S_3^- + e^{-i\theta_2}S_2^-S_3^+)  \\
 & & + (e^{-i(\theta_1+\theta_2)}S_3^+S_1^- + e^{i(\theta_1+\theta_2)}S_3^-S_1^+  
 ) \bigg\} \\
 & & +J_z^u (S_1^zS_2^z  +S_2^zS_3^z + S_3^zS_1^z),
\end{eqnarray*}
where three intra-spins are denoted by $\bm{S}_1,\bm{S}_2,\bm{S}_3$ and $S_k^+ = S_k^x+iS_k^y$ and 
$S_k^- = S_k^x-iS_k^y$ with $k=1,2,3$. We then define a Berry phase $\gamma_u(C)$ 
$$
 i\gamma_{u}(C) = \int_C \langle \psi (\Theta) | \bm{\nabla}_{\Theta} | \psi(\Theta)\rangle \cdot d\Theta
$$  
associated with the cyclic change of the 
parameter $\Theta$ along the path $C$ shown in Fig. \ref{fig1}(b).
Here $|\Psi(\Theta)\rangle$ represents the ground-state wave function for the parameter $\Theta$. 
As the path $C$ is equivalent to paths $C'$ and $C''$ because of the $Z_3$ symmetry of the Kagome lattice and $C + C' + C'' = 0$, we have $\gamma_u(C) = \gamma_u(C') = \gamma_u(C'')$ and $\gamma_u(C) + \gamma_u(C') + \gamma_u(C'') = 0$ mod $2\pi$. The Berry phase defined above is thus expected to be quantized to $\gamma_{u}(C) = 2\pi m/3$ with $m=0,1,2$ \cite{HM}. The quantization of the present $Z_3$ Berry phase is thus protected by the $Z_3$ symmetry of the Kagome lattice. Hereafter, we denote $\gamma_u(C)$ as $\gamma_u$. 

For the extreme case where the couplings between neighboring unit cells (thin lines in Fig. \ref{fig1}(a)) 
are negligibly small and the unit cells are decoupled from each other, the ground state is given by the product of
the ground states of the decoupled unit cells.
In such a case, the above modification of interactions reduces to the rotations of spins $(S_{1}^+,S_{3}^+) \to 
(S_{1}^+e^{i\theta_1}, S_{3}^+e^{-i\theta_2})$, which is 
a unitary transformation for the Hamiltonian in the space where the average magnetization is 1/3. \cite{HM} 
The ground state of the three spins in the unit cell   
is then given by 
$$
 |\psi(\Theta) \rangle = \frac{1}{\sqrt{3}} ( e^{i\theta_1}| \uparrow \downarrow \downarrow \rangle + 
 |  \downarrow \uparrow \downarrow \rangle + e^{-i\theta_2}|  \downarrow \downarrow\uparrow \rangle ),
$$
with the energy $(-J_{xy}-j_z^u/4)\hbar^2$, which is independent of $\Theta$.  Here, we assume that the magnitude of the average magnetization is 1/3 of the saturation magnetization in each 
unit cell. Subsequently, it is straightforward to see that the Berry phase defined above indeed  becomes $\gamma_{u} = 2\pi/3$. 
This quantized value of the Berry phase is broadly expected in the phases  
adiabatically connected to such an extreme case as well, namely, the tripartite entangled plaquette W state \cite{CCM}.

For a system with non-zero couplings between unit cells (thin lines),
we have performed exact diagonalization of Hamiltonians in finite systems to obtain the ground state numerically.
Using the ground states $|\psi(\Theta_n) \rangle$ for each value of the parameter $\Theta_n$ along the path $C$,
the Berry phase can be evaluated as\cite{FHS}
$
 \gamma_u = \arg (\prod_{n=0}^{N-1} \langle \psi(\Theta_{n+1} | \psi(\Theta_n) \rangle),
$
where $N$ denotes the number of mesh points along the path $C$. In practice, we set  
$\Theta_n = (2\pi/N)(n, n)$ for $0\leq n\leq N/3$ and $\Theta_n = (2\pi /N)(n, (N-n)/2)$ for $N/3 <  n \leq N$.
Note that the Berry phase can be defined as long as the gap above the ground state does not close along the path $C$. 

\begin{figure}
\begin{center}
\includegraphics[scale=0.45]{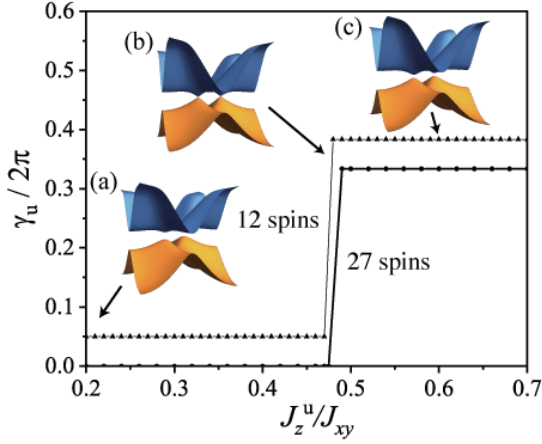}
\caption{(Color Online)
The evaluated Berry phase $\gamma_u$ for $J_z^d /J_{xy} = 0.4$ as a function of $J_z^u/J_{xy}$ for systems 
with 12 and 27 spins (Fig. \ref{fig1}(a)) with $N=15$. For 12 spins $\gamma_u +0.05$ is plotted instead of $\gamma_u$.
In both cases, it becomes 
$2\pi/3$ when the coupling $J_z^u/J_{xy}$ is larger than a critical value $0.475 \pm 0.01$. The energies of the ground state and 
the first excited states for 12 spins in the $\theta_1-\theta_2$ plane are also shown for $J_z^u/J_{xy}=$0.2(a), 0.475(b), and 0.6(c). 
The energy gap, which is symmetric with respect to the line $\theta_1+\theta_2 =2\pi$, closes only at the critical point (b) with $\theta_1=\theta_2=2\pi/3$ and $4\pi/3$.
\label{fig3}
}
\end{center}
\end{figure}

\begin{figure}
\begin{center}
\includegraphics[scale=0.3]{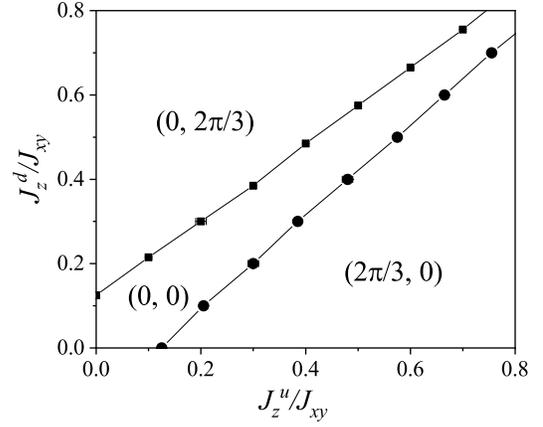}
\caption{
The phase boundary between $\gamma_{u(d)}=0$ and $\gamma_{u(d)} = 2\pi/3$ represented by 
the solid circles(squares) with lines obtained for the system of 27 spins (Fig. \ref{fig1}(a)) under the periodic boundary condition. 
Three phases are labeled by the Berry phases $(\gamma_u, \gamma_d)=(0,0), (2\pi/3,0)$ and $(0,2\pi/3)$.
\label{fig4}
}
\end{center}
\end{figure}

Adopting the Lanczos technique, we obtain the ground state
of the spin-1/2 XXZ model on the Kagome lattice up to 27 spins with the periodic boundary condition in the sub-space where the 
average magnetization is 1/3 of the saturation magnetization. We then find that the Berry phase 
is indeed quantized to an integer multiple of $2\pi/3$ and shows a transition from $\gamma_{u} = 0$ to 
$\gamma_{u} = 2\pi/3$ as the anisotropy $J_z^u/J_z^d$ is increased (Fig. \ref{fig3}).
This clearly indicates the ground state in a strongly anisotropic regime is adiabatically connected to 
the decoupled three-qubit W state. It should be noted that even in a small system with 27 spins, 
the signature of the non-trivial topological state can be captured by evaluating the Berry phase.  
We estimate the phase boundary between $\gamma_{u} = 0$ and $\gamma_{u} = 2\pi /3$ 
as shown in Fig. \ref{fig4}. We also evaluate the Berry phase $\gamma_d$ for the down triangles. 
The Berry phase $\gamma_d$ is also shown to have a quantized value $2\pi/3$ when the coupling $J_z^d/J_z^u$
becomes larger than a critical value (Fig. \ref{fig4}). As expected from the symmetry, the phase with $\gamma_u=2\pi/3$ and that with $\gamma_d=2\pi/3$
appear symmetrically with respect to the isotropic line $J_z^u = J_z^d$. The narrow region between these two phases with  
$\gamma_u = \gamma_d =0$ may correspond to the ferromagnetic phase.\cite{CCM}

In summary, we have investigated the Berry phase in a spin 1/2 XXZ model on the Kagome lattice when the average magnetization is 1/3
of the saturation magnetization. We have shown that the fractionally quantized $Z_3$ Berry phase becomes 
a quantized value $2\pi/3$ when the anisotropy is stronger than a critical value. 
This fact indicates the emergence of a topologically non-trivial phase and is 
consistent with the existence of the tripartite entangled plaquette phase in the anisotropic regime. 
Although the phase diagram shown in Fig. \ref{fig4} is qualitatively similar to that obtained by a large-scale Monte Carlo analysis\cite{CCM},
the critical values of $J_z^u/J_{xy}$ are smaller than those in Ref. \onlinecite{CCM}. This 
may be attributed to the small size of the systems in the present study, suggesting that further studies are necessary in much larger systems 
by adopting a Monte Carlo technique to 
evaluate the Berry phase \cite{MT} for quantitative studies of topological phases in two dimensions.

\begin{acknowledgments}

The authors thank H. Aoki for valuable communications. 
The work was supported in part by JSPS KAKENHI grant numbers 
JP15K05218 (TK), 
JP16K13845 (YH)
and JP17H06138. 

\end{acknowledgments}

\end{document}